\documentclass[%
superscriptaddress,
preprint,
 amsmath,amssymb,
aps,
]{revtex4-1}
\usepackage[dvipdfmx]{graphicx} 
\usepackage{bmpsize}
\usepackage{dcolumn}
\usepackage{bm}
\usepackage{natbib}
\usepackage{longtable}
\usepackage{adjustbox}
\usepackage{lipsum}
\usepackage{xcolor}

\makeatletter
\newcommand*{\rom}[1]{\expandafter\@slowromancap\romannumeral #1@}
\makeatother
\begin{document}


\title{Revealing and controlling nuclear dynamics following inner-shell photoionization of N$_2$}

\author{Qingli Jing}
\affiliation{School of Science, Jiangsu University of science of technology, 212003, Zhenjiang, Jiangsu, China}
\author{Hong Qian}
\affiliation{School of Science, Jiangsu University of science of technology, 212003, Zhenjiang, Jiangsu, China}
\author{Peng Xu}
\affiliation{Institute of Quantum Information and Technology, Nanjing University of Posts and Telecommunications, Nanjing, Jiangsu 210003, China}
\date{\today}%
\date{February 10, 2021}

\begin{abstract}
In this work, we apply the Monte Carlo wave packet method to study the ultrafast nuclear dynamics following inner-shell photoionization of N$_2$ exposed to an ultrashort intense X-ray pulse. The photon energy of the X-ray pulse is large enough to remove a $1s$ electron from the N atom in N$_2$. The intermediate state in N$_2^{+}$ is highly excited so that autoionization takes place from this state to the dissociative or non-dissociative electronic states of ungerade and gerade symmetries in N$_2^{++}$. The possible vibrational resonances allowed by the non-dissociative states prevents a direct extraction of the nuclear kinetic release (KER) spectrum from the nuclear wave packets in N$_2^{++}$. Therefore, we propose a hybrid technique by combining the advantages of two energy analysis strategies to obtain the final nuclear KER spectrum of the process. A femtosecond IR probe pulse, which couples the electronic states in N$_2^{++}$ together, is applied to achieve a time-resolved imaging and controlling of the ultrafast dynamics that takes place during double ionization of N$_2$. The influence of the laser parameters including the peak intensity, pulse duration and pump-probe delay, on the nuclear dynamics is also investigated. 
\end{abstract}

\maketitle
\section{INTRODUCTION} 
Study of ultrafast processes in molecules has attracted many research interests due to the potential applications in uncovering and controling of ultrafast electronic or nuclear dynamics in molecules with ultrashort and ultraintense light pulses. The phenomena related to the additional nuclear degrees of freedom in molecules have been investigated extensively, e.g., the laser-indced allignment and orientaion~\cite{allignPhysRevLett.102.023001,allignPhysRevA.70.063410,allignPhysRevA.100.043406}, rotational and vibrational autoionization~\cite{vibrationPhysRevLett.60.917,rotationPhysRevA.6.160,vibrationalautoionzation}, dissociation and dissocative ioinzation~\cite{dissociationPhysRevLett.77.4150,dissociationPhysRevLett.123.233202,dissociationPhysRevA.88.063420}, and charge-resonance-enhance ionization~\cite{CREIPhysRevA.52.R2511,CREIPhysRevA.59.2153,CREIPhysRevLett.107.063201}. Apart from that, the autoionization dynamics in molecules due to the electron correlation effect have been studied both theoretically and experimentally in the past decades, e.g, the Auger decay~\cite{ADPhysRevA.43.6053,ADPhysRevLett.105.213005,ADPhysRevLett.105.233001} and autoionzation from doubly excited states~\cite{AILafosse_2003,AIDPhysRevLett.110.213002}, by using the ultrashort XUV or X-ray pulses. \par

Considerable research works have been conducted by using ultrashort laser pulses to study the electron or nuclear dynamics during the ionization processes in molecules such as H$_2$~\cite{H2PhysRevLett.79.2022,H2PhysRevA.65.033403,H2Palacios_2015}, CO~\cite{COPhysRevLett.111.163001,COPhysRevA.101.013433}, N$_2$~\cite{N2PhysRevA.86.013415, N2PhysRevA.94.013426} and so on. For double or multiple ionization of molecules, the Coulomb-exposition imaging technique is extensively used to obtain the photoelectron kinetic energy spectrum or the nuclear KER spectrum of the charged fragments following dissociative ionization, from which the ultrafast electron or nuclear dynamics involved in each ionization channel can be identified and mapped out. From the theoretical point of view, a full quantum description of double ionization of a molecule, even for the simplest H$_2$, is an impossible task and approximations should be made to trace the nuclear dynamics involved. Methods based on the wave packet propagation method are capable to provide results that reproduces the main structures in the experimental counterpart.\par

There have been lots of research efforts put in studying the double ionization process of N$_2$~\cite{N2CPhysRevA.58.R4271,N2CPhysRevA.63.040701,N2CPhysRevLett.92.173001}. Compared with H$_2$, the dicatioic state of N$_2$ can support vibrational resonances where the N$_2^{++}$ stays bound and thus shows new characteristics for the nuclear dynamics. This work is mainly inspired by the earlier works of the molecular nitrogen~\cite{N2insPhysRevA.83.013417,N2insPhysRevA.86.043404,N2insPhysRevA.95.043419}. In Ref.~\cite{N2insPhysRevA.95.043419}, control over the Coulomb explosion imaging of N$_2^{++}$ is realized by using the X-ray-pump-IR probe setting. Apart from the Auger electron energy spectra and the yields of bound N$_2^{++}$ fragments at different Auger energy, we are more interested in recovering the nuclear dynamics involved by obtaining the nuclear KER spectra for different IR pulses with the Monte Carlo wave packet (MCWP) approach. This method has been proven to be especially useful in reproducing the nuclear KER spectra following double ionization of H$_2$~\cite{MCWPH2PhysRevA.81.053409,MCWPH2PhysRevA.81.053410,MCWPH2PhysRevA.94.063402,MCWPH2PhysRevA.97.043426} and O$_2$~\cite{MCWPO2PhysRevA.83.063415}. We will show later that our work can offer clear evidences for the laser-induced couplings between states in N$_2^{++}$, and thus provides a deeper insight of the dynamics involved in the process of interest. \par

This paper is organized as follows. We first provide a brief discussion of the process to be studied in Sec.~\ref{sec:process}. Then a summary of how to impliment the MCWP approach to simulate double ionization of N$_2$ is presented in Sec~\ref{sec:method}. We give in Sec.~\ref{sec:results} the simulation results of the nuclear KER spectra following autoionization of N$_2^{+}$ and provide detailed discussions for the nuclear dynamics involved . Section~\ref{sec:conclusion} concludes. Throughout this work, atomic units are used unless stated otherwise.
   
\section{\label{sec:process}Process}

\begin{figure}[!h]
\includegraphics[scale=0.9]{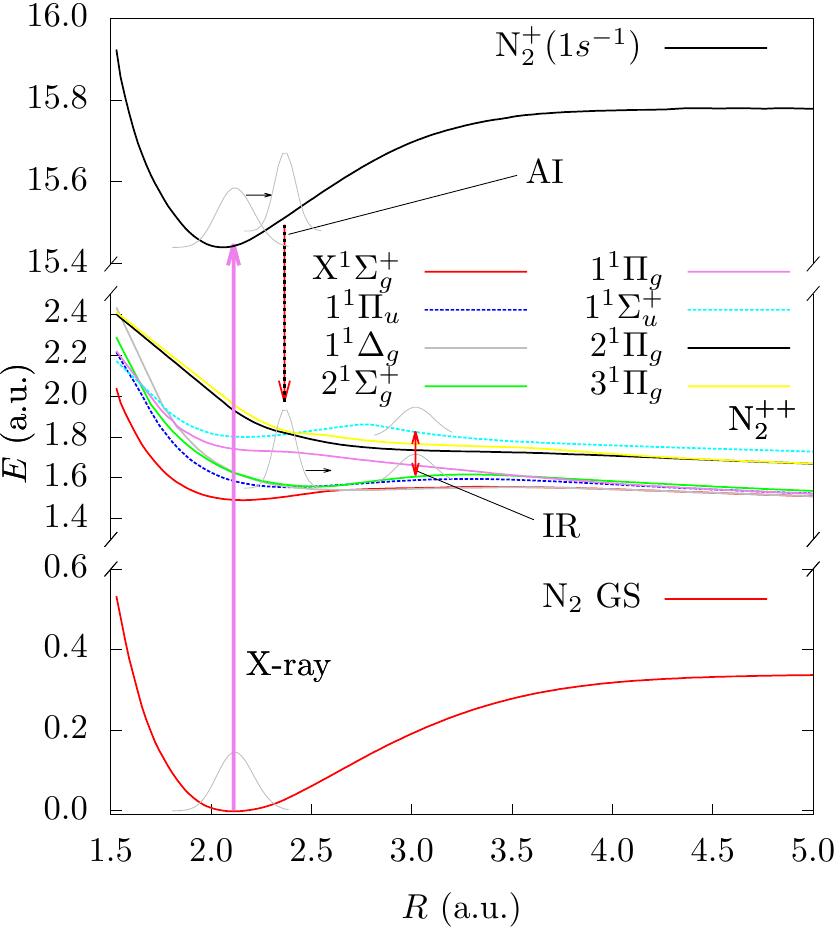}
\centering
\caption{Potential energy curves involved in the double ionization process for N$_2$ to be studied as a function of the internuclear distance $R$. From bottom to top at around $R=2.1$ a.u., the curves are for the ground state of N$_2$, the ground state and seven excited states of ungerade and gerade symmetries in N$_2^{++}$, and the intermediate state in N$_2^+(1s^{-1})$, respectively. These curves are taken from Ref.~\cite{N2insPhysRevA.95.043419}.}
\label{fig:potential_energy_curves}
\end{figure} 
Now we present a brief discussion for the process to be studied, i.e., the double ionization process of N$_2$. As shown in Fig.~\ref{fig:potential_energy_curves}, the N$_2$ molecule is initially in its ground electronic and vibrational state. Through interaction with an intense X-ray pulse, a $1s$ electron is removed from one of the nitrogen atom in N$_2$ and suddenly launches a nuclear wave packet evolving slowly in the inner-shell-ionized electronic state in N$_2^+$. Due to the electron correlation effect, subsequent autoionization from this unstable, highly excited state can take place, creating nuclear wave packets evolving along the potential energy curves for the electronic states in N$_2^{++}$. If a femtosecond IR pulse is then applied, the electronic states of gerade and ungerade symmetries in N$_2^{++}$ will interact with each other because of laser-induced dipole couplings between them. After the IR pulse, we can obtain the nuclear KER spectrum for double ionization of N$_2$ from the final nuclear wave packets in N$_2^{++}$,  which shows us the effect of the IR probe laser pulse on the nuclear dynamics in N$_2^{++}$. The nuclear KER spectra at different X-ray-pump-IR-probe delays are also capable of revealing the nuclear dynamics of the intermediate state in N$_2^+$.

\section{\label{sec:method}Method}  
In this work, we apply the MCWP approach to study the double ionization process of N$_2$ outlined in Sec.~\ref{sec:process}. The implementation and validation of this method in simulating double ionization of diatomic molecules have been discussed in detail elsewhere in Ref.~\cite{MCWPH2PhysRevA.81.053409,MCWPH2PhysRevA.81.053410,MCWPH2PhysRevA.94.063402,MCWPH2PhysRevA.97.043426}. To be summarized, this method adopts a non-Hermitian Hamiltonian to describe ionization of a molecular system as a decay process, i.e.,
\begin{equation}
H=H_{\text{s}}-\frac{i}{2}\sum_{mn}C_{mn}^+C_{mn}.
\label{eq:total Hamiltonian}
\end{equation}
The Hermitian system Hamiltonian $H_s$ in Eq.~(\ref{eq:total Hamiltonian}) is a superposition of three parts, the nuclear kinetic energy operator $T_N$, the field-free electronic Hamiltonian $H_{\text{e}}$ and the light-molecule interaction potential $V_I(t)=-\vec{\mu}\cdot\vec{F}(t)$ with $\mu$ the transition dipole moment operator and $\vec{F}(t)$ the electric field of the applied laser field, i.e., $H_s=T_N+H_{\text{e}}+V_I(t)$. $C_{mn}$ in the non-Hermitian term of Eq.~(\ref{eq:total Hamiltonian}) is a jump operator that specifies the transition pathway from the electronic state $|\phi_{{R},m}^{\text{e}}\rangle$ in the molecular system of concern to the electronic state $|\phi_{{R},n}^{\text{e}-1}\rangle$ in the system with one electron less, i.e., $C_{mn}=\int d\vec{R}\sqrt{\Gamma_{mn}(\vec{R},t)}|\phi_{{R},n}^{\text{e}-1}\rangle\langle \phi_{{R},m}^{\text{e}}|\otimes|\vec{R}\rangle\langle \vec{R}|$, where $\vec{R}$ is the position eigenket for the nuclear coordinate $\vec{R}$ and $\Gamma_{mn}(\vec{R},t)$ is the ionization rate from $|\phi_{{R},m}^{\text{e}}\rangle$ to state $|\phi_{{R},n}^{\text{e}-1}\rangle$.\par
To obtain the nuclear dynamics in a given molecular system, we solve the time-dependent Schr\"{o}dinger equation (TDSE) for the Hamiltonian in Eq.~(\ref{eq:total Hamiltonian}) by expressing the total wave function using the ansatz $|\Psi(t)\rangle=\int d\vec{R}\sum_m\chi(\vec{R},t)|\phi_{{R},m}^{\text{e}}\rangle\otimes|\vec{R}\rangle$. Here $|\phi_{{R},m}^{\text{e}}\rangle$ is the eigenstate satisfying the time-independent Schr\"odinger equation for the electronic Hamiltonian, i.e., $H_\text{e}|\phi_{{R},m}^{\text{e}}\rangle=E_m(\vec{R})|\phi_{{R},m}^{\text{e}}\rangle$, and $\chi(\vec{R},t)$ is the nuclear wave packet evolving along the potential energy surface $E_m(\vec{R})$ for the $|\phi_{{R},m}^{\text{e}}\rangle$ state. In the MCWP method, the total wave function of a molecule at a given charge state is a coherent superposition of the bound electronic states involved in the process. \par 
Taking the total wave function into the TDSE and projecting both sides of the TDSE on the eigenket $\langle\phi_{{R},m}^\text{e}|\langle\vec{R}|$, we obtain the evolution equation for the nuclear wave packet $\chi(\vec{R},t)$ in a given electronic state $|\phi_{{R},m}^{\text{e}}\rangle$
\begin{eqnarray}
i\dot\chi_m(\vec{R},t)&=&\big(T_{\text{N}}+E_m(\vec{R})-\frac{i}{2}\sum_n\Gamma_{mn}(\vec{R},t)\big)\chi_m(\vec{R},t) \nonumber \\ 
&+&\sum_{k\neq m}V_{mk}({\vec{R},t})\chi_k(\vec{R},t)
\label{eq:nuclear_wave_function_evolution}
\end{eqnarray}   
with $V_{mk}(\vec{R},t)=\langle\phi_{{R},m}^{\text{e}}|V_I(t)|\phi_{{R},k}^{\text{e}}\rangle=-\langle\phi_{{R},m}^{\text{e}}|\vec{\mu}|\phi_{{R},k}^{\text{e}}\rangle\cdot\vec{F}(t)=\vec{D}_{mk}(R)\cdot\vec{F}(t)$. $\vec{D}_{mk}(R)$ is the dipole moment function between the two states. When the molecule to be studied is assumed to be rotationally frozen, Eq.~(\ref{eq:nuclear_wave_function_evolution}) is reduced to a one-dimentional equaiton that is much easier to solve. This is a good approximation since the natural timescale for rotation of a molecule is usually much larger than that for vibration of the same molecule. Eq.~(\ref{eq:nuclear_wave_function_evolution}) can be solved by using the split-operator fast Fourier transform method~\cite{FFTFEIT1982412}. \par
When we apply the MCWP approach to simulate double ionization of N$_2$, we will solve Eq.~(\ref{eq:nuclear_wave_function_evolution}) in each charge state, i.e., N$_2$, N$_2^+$ and N$_2^{++}$, with the possible occurrence of quantum jumps. Such a full-time evolution of the nuclear wave packets in N$_2$ is a stochastic trajectory. In order to obtain physics results, we have to include such trajectories as many as possible. To reduce the computational costs, we will use the deterministic sampling method instead to pick up trajecoties with two quantum jumps, which was confirmed to perform well in earlier works~\cite{MCWPH2PhysRevA.81.053409,MCWPH2PhysRevA.81.053410,MCWPH2PhysRevA.94.063402,MCWPH2PhysRevA.97.043426}. In this case, the probability of each trajectory is related to the ionization probabilities at the first and second jump times, i.e., $t_1$ and $t_2$. The nuclear kinetic energy spectrum for a given trajectory is  
\begin{equation}
P_{E,mk}(t_2;t_1)=\left| \int dR K_E(R)\chi_m^{\text{N}_2^{++}}(R,t_e) \right|^2.
\label{eq:ker spectrum for a trajectories}
\end{equation}
We see from Fig.~\ref{fig:potential_energy_curves} that the potential energy curves for the electronic states in N$_2^{++}$ are not purely dissociative and thus some fragments can stay bound in N$_2^{++}$ and will not dissociate into N$^+$ and N$^+$. In Eq.~(\ref{eq:ker spectrum for a trajectories}), $K_E(R)$ is either the coulomb wave or the vibrational eigenstates for the electronic states in N$_2^+$. The nuclear kinetic energy spectrum following double ionization of N$_2$ is then obtained by a weighted sum of the KER spectra for the selected trajectories.
\begin{eqnarray}
P_E&= \sum\limits_{m,t_1,k,t_2}\,P_1(t_1)P_{1m}(t_1)P_2(t_2;t_1)P_{2k}(t_2;t_1)
\nonumber\\
& \times 
P_{E,mk}(t_2;t_1).
\label{eq:nuclear KER spectra}
\end{eqnarray} 
 \par
In this work, we consider the polarization axis of the linealy polarized laser pulse is parallel to the molecular axis. The simulation box is from $R_\text{min}=1.5$ to $R_\text{max}=45.4$ for 2048 $R$ grids and the time step is $\Delta t=1$. The other input data are takes as follows: the autoionization rates for the excited state in N$_2^+$, the potentail erengy curves for the electronic states involved, and the dipole moment functions $D_{mn}(R)$ between the electronic state are taken and extracted from Ref.~\cite{N2insPhysRevA.95.043419}. 
\section{\label{sec:results}RESULTS AND DISCUSSION}
\begin{figure}[!h]
\includegraphics[scale=0.9]{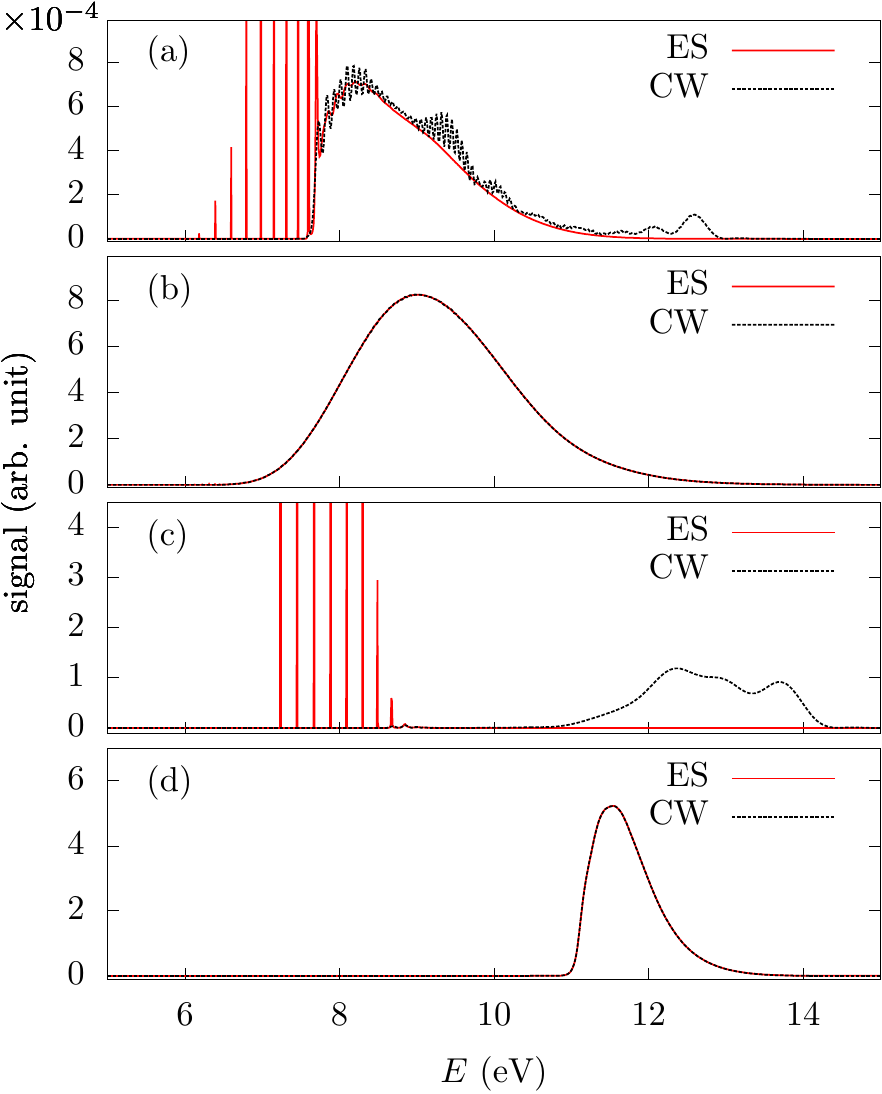}
\centering
\caption{Nuclear kinetic energy spectra following double ionization of N$_2$ interacting with the X-ray pulse only. (a), (b), (c), and (d) are results for the dominant autoionization pathways corresponding to the 2$^1\Sigma_g^+$, 1$^1\Delta_g$, 1$^1\Sigma_u^+$ and 2$^1\Pi_g$ states in N$_2^{++}$, respectively.}
\label{fig:fig2}
\end{figure} 
In this section, the simulation results from the MCWP method for the process considered in N$_2$ are presented. In Fig.~\ref{fig:fig2}, we plot the nuclear kinetic energy spectra for the dominant autoionization channels following double ionization of N$_2$ interacting with an X-ray pulse. In each panel, the solid line is for the result obtained by projecting the final nuclear wave function on the vibrational eigenstates of the potential energy curve (the first energy analysis method) and the dashed line for that by projecting the nuclear wave packet on the Coulomb waves for the $1/R$ curve (the second energy analysis method). When the potential energy curve is non-dissociative and deviates much from the $1/R$ curve, there is a large discrepancy between the two spectra. The first energy analysis method provides the accurate result, however, sharp resonances in the spectrum prevents us from obtaining the nuclear kinetic energy release spectra of N$_2^{++}$ that can be measured in the experiments. They correspond to the allowed vibrational states. Since all the potential energy curves in N$_2^{++}$ is closely parallel to the $1/R$ curve at large internuclear distances, the second energy analysis method gives very good results at small kinetic energies. Thus, we will combine the results from these two methods to get the nuclear KER spectrum contributed by a given state: for kinetic energies larger than 9 eV, apply the first method and it can provide results of the nuclear KER spectrum, and for kinetic energies smaller than 9 eV, the second energy analysis method is better to eliminate the sharp resonances.\par
\begin{figure}[!h]
\includegraphics[scale=1]{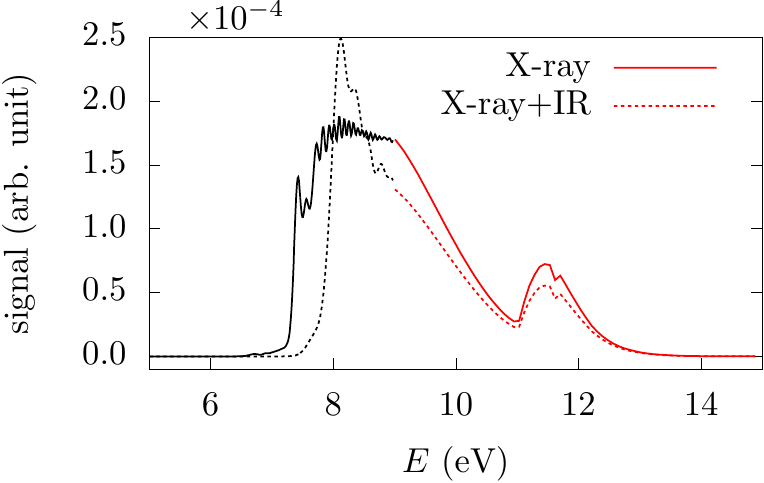}
\centering
\caption{Nucler KER spectra following double ionization of N$_2$ interacting with the X-ray pulse only and with an IR probe pulse with zero delay. The parameters of the IR laser are as follows: the peak intensity is $I_0=2.24\times10^{14}$, the pulse duration is 4 optical cycles, the wavelength is 800 nm and it has a zero delay with respect to the X-ray pulse.}
\label{fig:fig3}
\end{figure} 
Here is the nuclear KER spectra following double ionization of N$_2$ interacting with an X-ray pulse obtained by using the above-mentioned techniques, as shown in Fig.~\ref{fig:fig3}. We also provide the nuclear KER spectrum for N$_2^{++}$ interacting with an IR probe pulse for comparison. The parameters for the applied Gaussian-shape IR pulse are as follows: the pump-probe delay is $\tau=$0, the wavelength is 800 nm, the peak intensity is $2.24\times10^{14}$ W$/$cm$_2$ and the FWHM duration is 4 optical cycles. The overall shapes of the two spectrum with or without the IR pulse both peak at around $E=8$ eV and $E=11$ eV. The signal around the high energy peak is mainly from autoionization to dissociative states in N$_2^{++}$ and the signal around the low energy peak mainly results from tunnel dissociation at the non-dissociative states. The main difference between the two spectra is that for the spectrum with an IR probe pulse that a small part of the signal around the KER value $E=11$ eV is transferred to the signal around $E=8$ eV.  \par
\begin{figure}[!h]
\includegraphics[scale=0.9]{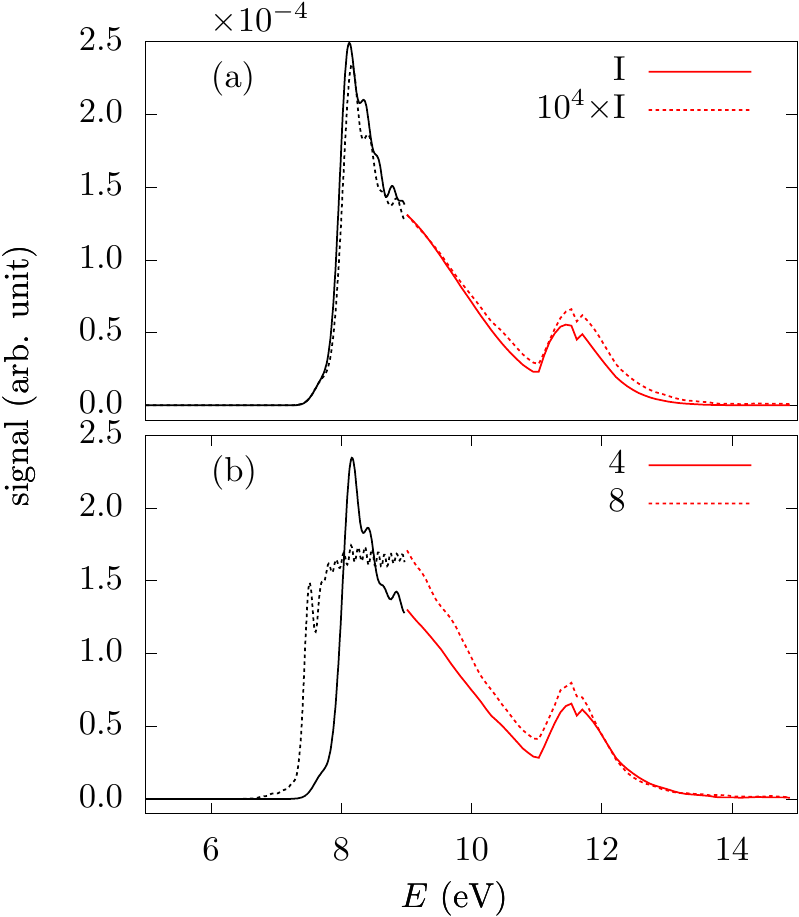}
\centering
\caption{(a) Nuclear KER spectra following double ionization of N$_2$ interacting with IR pulses at peak intensities of $I_0=2.24\times10^{14}$ and  $I_0=2.24\times10^{18}$ W$/$cm$^2$. (b) Nuclear KER spectra following double ionization of N$_2$ interacting with IR pulses with pulses duration of 4 and 8 optical cycles. The other parameters for the IR pulses are identical to the IR pulse in Fig.~\ref{fig:fig3}.}
\label{fig:fig4}
\end{figure} 
Now let us explore the influence of the laser parameters of the IR pulse such as the peak intensity and the pulse duration on the nuclear KER spectrum. We present the corresponding results in Fig.~\ref{fig:fig4}. Figure~\ref{fig:fig4}(a) illustrates the nuclear KER spectra for N$_2^{++}$ interacting with two peak intensities, $I_0=2.24\times10^{14}$ and  $I_0=2.24\times10^{18}$ W$/$cm$^2$. The two spectrum looks very similar except that for the larger intensity case the signal around the KER value $E=11$ eV is enhanced while the signal around $E=8$ eV is suppressed, indicating that when the IR field becomes stronger, the IR pulse induces more transitions from the non-dissociative states to the dissociative states in N$_2^{++}$. In Fig.~\ref{fig:fig4}(b), the nuclear KER spectra for N$_2^{++}$ at two pulse durations, i.e., 4 and 8 optical cycles, are plotted. Just as expected, we observe a similar trend as the former case since a longer pulse means a longer time for dipole couplings between the states in N$_2^{++}$ to take place.  \par

\begin{figure}[!h]
\includegraphics[scale=1]{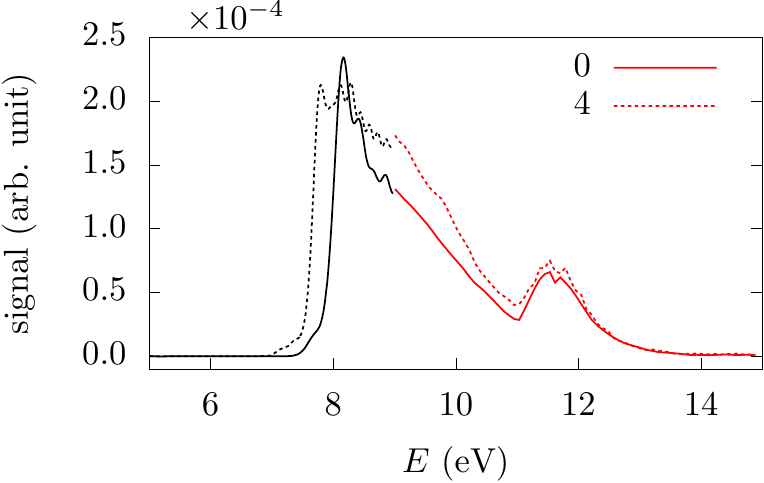}
\centering
\caption{Nuclear KER spectra following double ionization of N$_2$ interacting with IR pulses at different pump-probe delays, i.e., 0 and 4 optical cycles. The other laser parameters are identical to the IR pulse with a larger intensity in Fig.~\ref{fig:fig4}(a). }
\label{fig:fig5}
\end{figure}
We are also interested in investigating the influence of the pump-probe delay on the final nuclear KER spectrum. Figure~\ref{fig:fig5} shows the spectrum for N$_2^{++}$ interacting with IR laser pulses at different delays, i.e., zero delay and 4 optical cycles. We see that for a larger delay, the spectrum moves towards lower kinetic energies. This means that the nuclear wave packet in the highly excited state in N$_2^{+}$ moves towards a larger internuclear distance during the delay time. In fact, a more detail analysis of the structure of the time-resolved nuclear KER spectra can also help to recover the nuclear dynamics in N$_2^{++}$.  

\section{\label{sec:conclusion}CONCLUSION}
In this work, we investigate the ultrafast nuclear dynamics following inner-shell photoionization of N$_2$ interacting with an X-ray pulse by using the MCWP method. We have obtained the nuclear KER spectrum of N$_2^{++}$ from the non-dissociative states that support sharp vibrational resonances by applying a hybrid energy analysis method. The influences of the parameters of the IR probe pulses which couples the states in N$_2^{++}$ together was also investigated to show the possibility of controlling ultrafast nuclear dynamics by using IR pulses. From the obtained nuclear KER spectra, we can see that the intensity and pulse duration of the IR pulse have similar effect on the nuclear dynamics.  When the IR pulse is stronger or longer, there is a larger KER signal at larger kinetic energies due to laser-induced couplings taking place at smaller internuclear distances. Apart from that, we find that the final nuclear KER spectrum moves to smaller kinetic energies for larger pump-probe delays, which can reflect the nuclear dynamics involved in the highly excited states in N$_2^+$. 
\begin{acknowledgments}
This work was financially supported by the National Natural Science Foundation of China under Grant No. 2055011906. 
Q. Jing was also supported by the Scientific Research Foundation of Jiangsu University of Science and Technology (Grant No. 1052931901) and Jiangsu Province Innovative and Entrepreneurial Doctiral Fund 2020 (Grant No. 1054902004).
P. Xu was supported by the Scientific Research Foundation of Nanjing University of Posts and Telecommunications (Grant No. NY218097) and the Young Fund of Jiangsu Natural Science Foundation of China (Grant No. BK20180750).
We also acknowledge support from the Young Fund of Jiangsu Natural Science Foundation of China (Grant No. BK20190953 and No. BK20190954).
\end{acknowledgments}



\begin{thebibliography}{36}%
\makeatletter
\providecommand \@ifxundefined [1]{%
 \@ifx{#1\undefined}
}%
\providecommand \@ifnum [1]{%
 \ifnum #1\expandafter \@firstoftwo
 \else \expandafter \@secondoftwo
 \fi
}%
\providecommand \@ifx [1]{%
 \ifx #1\expandafter \@firstoftwo
 \else \expandafter \@secondoftwo
 \fi
}%
\providecommand \natexlab [1]{#1}%
\providecommand \enquote  [1]{``#1''}%
\providecommand \bibnamefont  [1]{#1}%
\providecommand \bibfnamefont [1]{#1}%
\providecommand \citenamefont [1]{#1}%
\providecommand \href@noop [0]{\@secondoftwo}%
\providecommand \href [0]{\begingroup \@sanitize@url \@href}%
\providecommand \@href[1]{\@@startlink{#1}\@@href}%
\providecommand \@@href[1]{\endgroup#1\@@endlink}%
\providecommand \@sanitize@url [0]{\catcode `\\12\catcode `\$12\catcode
  `\&12\catcode `\#12\catcode `\^12\catcode `\_12\catcode `\%12\relax}%
\providecommand \@@startlink[1]{}%
\providecommand \@@endlink[0]{}%
\providecommand \url  [0]{\begingroup\@sanitize@url \@url }%
\providecommand \@url [1]{\endgroup\@href {#1}{\urlprefix }}%
\providecommand \urlprefix  [0]{URL }%
\providecommand \Eprint [0]{\href }%
\providecommand \doibase [0]{http://dx.doi.org/}%
\providecommand \selectlanguage [0]{\@gobble}%
\providecommand \bibinfo  [0]{\@secondoftwo}%
\providecommand \bibfield  [0]{\@secondoftwo}%
\providecommand \translation [1]{[#1]}%
\providecommand \BibitemOpen [0]{}%
\providecommand \bibitemStop [0]{}%
\providecommand \bibitemNoStop [0]{.\EOS\space}%
\providecommand \EOS [0]{\spacefactor3000\relax}%
\providecommand \BibitemShut  [1]{\csname bibitem#1\endcsname}%
\let\auto@bib@innerbib\@empty
\bibitem [{\citenamefont {Holmegaard}\ \emph {et~al.}(2009)\citenamefont
  {Holmegaard}, \citenamefont {Nielsen}, \citenamefont {Nevo}, \citenamefont
  {Stapelfeldt}, \citenamefont {Filsinger}, \citenamefont {K\"upper},\ and\
  \citenamefont {Meijer}}]{allignPhysRevLett.102.023001}%
  \BibitemOpen
  \bibfield  {author} {\bibinfo {author} {\bibfnamefont {L.}~\bibnamefont
  {Holmegaard}}, \bibinfo {author} {\bibfnamefont {J.~H.}\ \bibnamefont
  {Nielsen}}, \bibinfo {author} {\bibfnamefont {I.}~\bibnamefont {Nevo}},
  \bibinfo {author} {\bibfnamefont {H.}~\bibnamefont {Stapelfeldt}}, \bibinfo
  {author} {\bibfnamefont {F.}~\bibnamefont {Filsinger}}, \bibinfo {author}
  {\bibfnamefont {J.}~\bibnamefont {K\"upper}}, \ and\ \bibinfo {author}
  {\bibfnamefont {G.}~\bibnamefont {Meijer}},\ }\bibfield  {title} {\enquote
  {\bibinfo {title} {Laser-induced alignment and orientation of
  quantum-state-selected large molecules},}\ }\href {\doibase
  10.1103/PhysRevLett.102.023001} {\bibfield  {journal} {\bibinfo  {journal}
  {Phys. Rev. Lett.}\ }\textbf {\bibinfo {volume} {102}},\ \bibinfo {pages}
  {023001} (\bibinfo {year} {2009})}\BibitemShut {NoStop}%
\bibitem [{\citenamefont {P\'eronne}\ \emph {et~al.}(2004)\citenamefont
  {P\'eronne}, \citenamefont {Poulsen}, \citenamefont {Stapelfeldt},
  \citenamefont {Bisgaard}, \citenamefont {Hamilton},\ and\ \citenamefont
  {Seideman}}]{allignPhysRevA.70.063410}%
  \BibitemOpen
  \bibfield  {author} {\bibinfo {author} {\bibfnamefont {E.}~\bibnamefont
  {P\'eronne}}, \bibinfo {author} {\bibfnamefont {M.~D.}\ \bibnamefont
  {Poulsen}}, \bibinfo {author} {\bibfnamefont {H.}~\bibnamefont
  {Stapelfeldt}}, \bibinfo {author} {\bibfnamefont {C.~Z.}\ \bibnamefont
  {Bisgaard}}, \bibinfo {author} {\bibfnamefont {E.}~\bibnamefont {Hamilton}},
  \ and\ \bibinfo {author} {\bibfnamefont {T.}~\bibnamefont {Seideman}},\
  }\bibfield  {title} {\enquote {\bibinfo {title} {Nonadiabatic laser-induced
  alignment of iodobenzene molecules},}\ }\href {\doibase
  10.1103/PhysRevA.70.063410} {\bibfield  {journal} {\bibinfo  {journal} {Phys.
  Rev. A}\ }\textbf {\bibinfo {volume} {70}},\ \bibinfo {pages} {063410}
  (\bibinfo {year} {2004})}\BibitemShut {NoStop}%
\bibitem [{\citenamefont {Tutunnikov}\ \emph {et~al.}(2019)\citenamefont
  {Tutunnikov}, \citenamefont {Flo\ss{}}, \citenamefont {Gershnabel},
  \citenamefont {Brumer},\ and\ \citenamefont
  {Averbukh}}]{allignPhysRevA.100.043406}%
  \BibitemOpen
  \bibfield  {author} {\bibinfo {author} {\bibfnamefont {I.}~\bibnamefont
  {Tutunnikov}}, \bibinfo {author} {\bibfnamefont {J.}~\bibnamefont
  {Flo\ss{}}}, \bibinfo {author} {\bibfnamefont {E.}~\bibnamefont
  {Gershnabel}}, \bibinfo {author} {\bibfnamefont {P.}~\bibnamefont {Brumer}},
  \ and\ \bibinfo {author} {\bibfnamefont {I.~S.}\ \bibnamefont {Averbukh}},\
  }\bibfield  {title} {\enquote {\bibinfo {title} {Laser-induced persistent
  orientation of chiral molecules},}\ }\href {\doibase
  10.1103/PhysRevA.100.043406} {\bibfield  {journal} {\bibinfo  {journal}
  {Phys. Rev. A}\ }\textbf {\bibinfo {volume} {100}},\ \bibinfo {pages}
  {043406} (\bibinfo {year} {2019})}\BibitemShut {NoStop}%
\bibitem [{\citenamefont {Bordas}\ \emph {et~al.}(1988)\citenamefont {Bordas},
  \citenamefont {Brevet}, \citenamefont {Broyer}, \citenamefont {Chevaleyre},
  \citenamefont {Labastie},\ and\ \citenamefont
  {Perrot}}]{vibrationPhysRevLett.60.917}%
  \BibitemOpen
  \bibfield  {author} {\bibinfo {author} {\bibfnamefont {C.}~\bibnamefont
  {Bordas}}, \bibinfo {author} {\bibfnamefont {P.~F.}\ \bibnamefont {Brevet}},
  \bibinfo {author} {\bibfnamefont {M.}~\bibnamefont {Broyer}}, \bibinfo
  {author} {\bibfnamefont {J.}~\bibnamefont {Chevaleyre}}, \bibinfo {author}
  {\bibfnamefont {P.}~\bibnamefont {Labastie}}, \ and\ \bibinfo {author}
  {\bibfnamefont {J.~P.}\ \bibnamefont {Perrot}},\ }\bibfield  {title}
  {\enquote {\bibinfo {title} {Electric-field--hindered vibrational
  autoionization in molecular {R}ydberg states},}\ }\href {\doibase
  10.1103/PhysRevLett.60.917} {\bibfield  {journal} {\bibinfo  {journal} {Phys.
  Rev. Lett.}\ }\textbf {\bibinfo {volume} {60}},\ \bibinfo {pages} {917--920}
  (\bibinfo {year} {1988})}\BibitemShut {NoStop}%
\bibitem [{\citenamefont {Dill}(1972)}]{rotationPhysRevA.6.160}%
  \BibitemOpen
  \bibfield  {author} {\bibinfo {author} {\bibfnamefont {D.}~\bibnamefont
  {Dill}},\ }\bibfield  {title} {\enquote {\bibinfo {title} {Angular
  distributions of photoelectrons from {H}${_{2}}$: Effects of rotational
  autoionization},}\ }\href {\doibase 10.1103/PhysRevA.6.160} {\bibfield
  {journal} {\bibinfo  {journal} {Phys. Rev. A}\ }\textbf {\bibinfo {volume}
  {6}},\ \bibinfo {pages} {160--172} (\bibinfo {year} {1972})}\BibitemShut
  {NoStop}%
\bibitem [{\citenamefont {Dehmer}\ and\ \citenamefont
  {Chupka}(1977)}]{vibrationalautoionzation}%
  \BibitemOpen
  \bibfield  {author} {\bibinfo {author} {\bibfnamefont {P.~M.}\ \bibnamefont
  {Dehmer}}\ and\ \bibinfo {author} {\bibfnamefont {W.~A.}\ \bibnamefont
  {Chupka}},\ }\bibfield  {title} {\enquote {\bibinfo {title} {On the mechanism
  for vibrational autoionization in {H}$_2$ },}\ }\href {\doibase
  10.1063/1.434154} {\bibfield  {journal} {\bibinfo  {journal} {J. Chem.
  Phys.}\ }\textbf {\bibinfo {volume} {66}},\ \bibinfo {pages} {1972--1981}
  (\bibinfo {year} {1977})}\BibitemShut {NoStop}%
\bibitem [{\citenamefont {Dietrich}\ \emph {et~al.}(1996)\citenamefont
  {Dietrich}, \citenamefont {Ivanov}, \citenamefont {Ilkov},\ and\
  \citenamefont {Corkum}}]{dissociationPhysRevLett.77.4150}%
  \BibitemOpen
  \bibfield  {author} {\bibinfo {author} {\bibfnamefont {P.}~\bibnamefont
  {Dietrich}}, \bibinfo {author} {\bibfnamefont {M.~Y.}\ \bibnamefont
  {Ivanov}}, \bibinfo {author} {\bibfnamefont {F.~A.}\ \bibnamefont {Ilkov}}, \
  and\ \bibinfo {author} {\bibfnamefont {P.~B.}\ \bibnamefont {Corkum}},\
  }\bibfield  {title} {\enquote {\bibinfo {title} {Two-electron dissociative
  ionization of {H}$_{2}$ and {D}$_{2}$ in infrared laser fields},}\ }\href
  {\doibase 10.1103/PhysRevLett.77.4150} {\bibfield  {journal} {\bibinfo
  {journal} {Phys. Rev. Lett.}\ }\textbf {\bibinfo {volume} {77}},\ \bibinfo
  {pages} {4150--4153} (\bibinfo {year} {1996})}\BibitemShut {NoStop}%
\bibitem [{\citenamefont {Ji}\ \emph {et~al.}(2019)\citenamefont {Ji},
  \citenamefont {Pan}, \citenamefont {He}, \citenamefont {Wang}, \citenamefont
  {Lu}, \citenamefont {Li}, \citenamefont {Gong}, \citenamefont {Lin},
  \citenamefont {Zhang}, \citenamefont {Ma}, \citenamefont {Li}, \citenamefont
  {Duan}, \citenamefont {Liu}, \citenamefont {Bai}, \citenamefont {Li},
  \citenamefont {He},\ and\ \citenamefont
  {Wu}}]{dissociationPhysRevLett.123.233202}%
  \BibitemOpen
  \bibfield  {author} {\bibinfo {author} {\bibfnamefont {Q.}~\bibnamefont
  {Ji}}, \bibinfo {author} {\bibfnamefont {S.}~\bibnamefont {Pan}}, \bibinfo
  {author} {\bibfnamefont {P.}~\bibnamefont {He}}, \bibinfo {author}
  {\bibfnamefont {J.}~\bibnamefont {Wang}}, \bibinfo {author} {\bibfnamefont
  {P.}~\bibnamefont {Lu}}, \bibinfo {author} {\bibfnamefont {H.}~\bibnamefont
  {Li}}, \bibinfo {author} {\bibfnamefont {X.}~\bibnamefont {Gong}}, \bibinfo
  {author} {\bibfnamefont {K.}~\bibnamefont {Lin}}, \bibinfo {author}
  {\bibfnamefont {W.}~\bibnamefont {Zhang}}, \bibinfo {author} {\bibfnamefont
  {J.}~\bibnamefont {Ma}}, \bibinfo {author} {\bibfnamefont {H.}~\bibnamefont
  {Li}}, \bibinfo {author} {\bibfnamefont {C.}~\bibnamefont {Duan}}, \bibinfo
  {author} {\bibfnamefont {P.}~\bibnamefont {Liu}}, \bibinfo {author}
  {\bibfnamefont {Y.}~\bibnamefont {Bai}}, \bibinfo {author} {\bibfnamefont
  {R.}~\bibnamefont {Li}}, \bibinfo {author} {\bibfnamefont {F.}~\bibnamefont
  {He}}, \ and\ \bibinfo {author} {\bibfnamefont {J.}~\bibnamefont {Wu}},\
  }\bibfield  {title} {\enquote {\bibinfo {title} {Timing dissociative
  ionization of {H}$_{2}$ using a polarization-skewed femtosecond laser
  pulse},}\ }\href {\doibase 10.1103/PhysRevLett.123.233202} {\bibfield
  {journal} {\bibinfo  {journal} {Phys. Rev. Lett.}\ }\textbf {\bibinfo
  {volume} {123}},\ \bibinfo {pages} {233202} (\bibinfo {year}
  {2019})}\BibitemShut {NoStop}%
\bibitem [{\citenamefont {Yue}\ and\ \citenamefont
  {Madsen}(2013)}]{dissociationPhysRevA.88.063420}%
  \BibitemOpen
  \bibfield  {author} {\bibinfo {author} {\bibfnamefont {L.}~\bibnamefont
  {Yue}}\ and\ \bibinfo {author} {\bibfnamefont {L.~B.}\ \bibnamefont
  {Madsen}},\ }\bibfield  {title} {\enquote {\bibinfo {title} {Dissociation and
  dissociative ionization of {H}${}_{2}{}^{+}$ using the time-dependent surface
  flux method},}\ }\href {\doibase 10.1103/PhysRevA.88.063420} {\bibfield
  {journal} {\bibinfo  {journal} {Phys. Rev. A}\ }\textbf {\bibinfo {volume}
  {88}},\ \bibinfo {pages} {063420} (\bibinfo {year} {2013})}\BibitemShut
  {NoStop}%
\bibitem [{\citenamefont {Zuo}\ and\ \citenamefont
  {Bandrauk}(1995)}]{CREIPhysRevA.52.R2511}%
  \BibitemOpen
  \bibfield  {author} {\bibinfo {author} {\bibfnamefont {T.}~\bibnamefont
  {Zuo}}\ and\ \bibinfo {author} {\bibfnamefont {A.~D.}\ \bibnamefont
  {Bandrauk}},\ }\bibfield  {title} {\enquote {\bibinfo {title}
  {Charge-resonance-enhanced ionization of diatomic molecular ions by intense
  lasers},}\ }\href {\doibase 10.1103/PhysRevA.52.R2511} {\bibfield  {journal}
  {\bibinfo  {journal} {Phys. Rev. A}\ }\textbf {\bibinfo {volume} {52}},\
  \bibinfo {pages} {R2511--R2514} (\bibinfo {year} {1995})}\BibitemShut
  {NoStop}%
\bibitem [{\citenamefont {Bandrauk}\ and\ \citenamefont
  {Ruel}(1999)}]{CREIPhysRevA.59.2153}%
  \BibitemOpen
  \bibfield  {author} {\bibinfo {author} {\bibfnamefont {A.~D.}\ \bibnamefont
  {Bandrauk}}\ and\ \bibinfo {author} {\bibfnamefont {J.}~\bibnamefont
  {Ruel}},\ }\bibfield  {title} {\enquote {\bibinfo {title}
  {Charge-resonance-enhanced ionization of molecular ions in intense laser
  pulses: Geometric and orientation effects},}\ }\href {\doibase
  10.1103/PhysRevA.59.2153} {\bibfield  {journal} {\bibinfo  {journal} {Phys.
  Rev. A}\ }\textbf {\bibinfo {volume} {59}},\ \bibinfo {pages} {2153--2162}
  (\bibinfo {year} {1999})}\BibitemShut {NoStop}%
\bibitem [{\citenamefont {Bocharova}\ \emph
  {et~al.}(2011{\natexlab{a}})\citenamefont {Bocharova}, \citenamefont
  {Karimi}, \citenamefont {Penka}, \citenamefont {Brichta}, \citenamefont
  {Lassonde}, \citenamefont {Fu}, \citenamefont {Kieffer}, \citenamefont
  {Bandrauk}, \citenamefont {Litvinyuk}, \citenamefont {Sanderson},\ and\
  \citenamefont {L\'egar\'e}}]{CREIPhysRevLett.107.063201}%
  \BibitemOpen
  \bibfield  {author} {\bibinfo {author} {\bibfnamefont {I.}~\bibnamefont
  {Bocharova}}, \bibinfo {author} {\bibfnamefont {R.}~\bibnamefont {Karimi}},
  \bibinfo {author} {\bibfnamefont {E.~F.}\ \bibnamefont {Penka}}, \bibinfo
  {author} {\bibfnamefont {J.-P.}\ \bibnamefont {Brichta}}, \bibinfo {author}
  {\bibfnamefont {P.}~\bibnamefont {Lassonde}}, \bibinfo {author}
  {\bibfnamefont {X.}~\bibnamefont {Fu}}, \bibinfo {author} {\bibfnamefont
  {J.-C.}\ \bibnamefont {Kieffer}}, \bibinfo {author} {\bibfnamefont {A.~D.}\
  \bibnamefont {Bandrauk}}, \bibinfo {author} {\bibfnamefont {I.}~\bibnamefont
  {Litvinyuk}}, \bibinfo {author} {\bibfnamefont {J.}~\bibnamefont
  {Sanderson}}, \ and\ \bibinfo {author} {\bibfnamefont {F.~m.~c.}\
  \bibnamefont {L\'egar\'e}},\ }\bibfield  {title} {\enquote {\bibinfo {title}
  {Charge resonance enhanced ionization of {CO}$_{2}$ probed by laser coulomb
  explosion imaging},}\ }\href {\doibase 10.1103/PhysRevLett.107.063201}
  {\bibfield  {journal} {\bibinfo  {journal} {Phys. Rev. Lett.}\ }\textbf
  {\bibinfo {volume} {107}},\ \bibinfo {pages} {063201} (\bibinfo {year}
  {2011}{\natexlab{a}})}\BibitemShut {NoStop}%
\bibitem [{\citenamefont {Coville}\ and\ \citenamefont
  {Thomas}(1991)}]{ADPhysRevA.43.6053}%
  \BibitemOpen
  \bibfield  {author} {\bibinfo {author} {\bibfnamefont {M.}~\bibnamefont
  {Coville}}\ and\ \bibinfo {author} {\bibfnamefont {T.~D.}\ \bibnamefont
  {Thomas}},\ }\bibfield  {title} {\enquote {\bibinfo {title} {Molecular
  effects on inner-shell lifetimes: Possible test of the one-center model of
  {A}uger decay},}\ }\href {\doibase 10.1103/PhysRevA.43.6053} {\bibfield
  {journal} {\bibinfo  {journal} {Phys. Rev. A}\ }\textbf {\bibinfo {volume}
  {43}},\ \bibinfo {pages} {6053--6056} (\bibinfo {year} {1991})}\BibitemShut
  {NoStop}%
\bibitem [{\citenamefont {Eland}\ \emph {et~al.}(2010)\citenamefont {Eland},
  \citenamefont {Tashiro}, \citenamefont {Linusson}, \citenamefont {Ehara},
  \citenamefont {Ueda},\ and\ \citenamefont
  {Feifel}}]{ADPhysRevLett.105.213005}%
  \BibitemOpen
  \bibfield  {author} {\bibinfo {author} {\bibfnamefont {J.~H.~D.}\
  \bibnamefont {Eland}}, \bibinfo {author} {\bibfnamefont {M.}~\bibnamefont
  {Tashiro}}, \bibinfo {author} {\bibfnamefont {P.}~\bibnamefont {Linusson}},
  \bibinfo {author} {\bibfnamefont {M.}~\bibnamefont {Ehara}}, \bibinfo
  {author} {\bibfnamefont {K.}~\bibnamefont {Ueda}}, \ and\ \bibinfo {author}
  {\bibfnamefont {R.}~\bibnamefont {Feifel}},\ }\bibfield  {title} {\enquote
  {\bibinfo {title} {Double core hole creation and subsequent {A}uger decay in
  {NH}$_{3}$ and {CH}$_{4}$ molecules},}\ }\href {\doibase
  10.1103/PhysRevLett.105.213005} {\bibfield  {journal} {\bibinfo  {journal}
  {Phys. Rev. Lett.}\ }\textbf {\bibinfo {volume} {105}},\ \bibinfo {pages}
  {213005} (\bibinfo {year} {2010})}\BibitemShut {NoStop}%
\bibitem [{\citenamefont {Travnikova}\ \emph {et~al.}(2010)\citenamefont
  {Travnikova}, \citenamefont {Liu}, \citenamefont {Lindblad}, \citenamefont
  {Nicolas}, \citenamefont {S\"oderstr\"om}, \citenamefont {Kimberg},
  \citenamefont {Gel'mukhanov},\ and\ \citenamefont
  {Miron}}]{ADPhysRevLett.105.233001}%
  \BibitemOpen
  \bibfield  {author} {\bibinfo {author} {\bibfnamefont {O.}~\bibnamefont
  {Travnikova}}, \bibinfo {author} {\bibfnamefont {J.-C.}\ \bibnamefont {Liu}},
  \bibinfo {author} {\bibfnamefont {A.}~\bibnamefont {Lindblad}}, \bibinfo
  {author} {\bibfnamefont {C.}~\bibnamefont {Nicolas}}, \bibinfo {author}
  {\bibfnamefont {J.}~\bibnamefont {S\"oderstr\"om}}, \bibinfo {author}
  {\bibfnamefont {V.}~\bibnamefont {Kimberg}}, \bibinfo {author} {\bibfnamefont
  {F.}~\bibnamefont {Gel'mukhanov}}, \ and\ \bibinfo {author} {\bibfnamefont
  {C.}~\bibnamefont {Miron}},\ }\bibfield  {title} {\enquote {\bibinfo {title}
  {Circularly polarized {X} rays: Another probe of ultrafast molecular decay
  dynamics},}\ }\href {\doibase 10.1103/PhysRevLett.105.233001} {\bibfield
  {journal} {\bibinfo  {journal} {Phys. Rev. Lett.}\ }\textbf {\bibinfo
  {volume} {105}},\ \bibinfo {pages} {233001} (\bibinfo {year}
  {2010})}\BibitemShut {NoStop}%
\bibitem [{\citenamefont {Lafosse}\ \emph {et~al.}(2003)\citenamefont
  {Lafosse}, \citenamefont {Lebech}, \citenamefont {Brenot}, \citenamefont
  {Guyon}, \citenamefont {Spielberger}, \citenamefont {Jagutzki}, \citenamefont
  {Houver},\ and\ \citenamefont {Dowek}}]{AILafosse_2003}%
  \BibitemOpen
  \bibfield  {author} {\bibinfo {author} {\bibfnamefont {A.}~\bibnamefont
  {Lafosse}}, \bibinfo {author} {\bibfnamefont {M.}~\bibnamefont {Lebech}},
  \bibinfo {author} {\bibfnamefont {J.~C.}\ \bibnamefont {Brenot}}, \bibinfo
  {author} {\bibfnamefont {P.~M.}\ \bibnamefont {Guyon}}, \bibinfo {author}
  {\bibfnamefont {L.}~\bibnamefont {Spielberger}}, \bibinfo {author}
  {\bibfnamefont {O.}~\bibnamefont {Jagutzki}}, \bibinfo {author}
  {\bibfnamefont {J.~C.}\ \bibnamefont {Houver}}, \ and\ \bibinfo {author}
  {\bibfnamefont {D.}~\bibnamefont {Dowek}},\ }\bibfield  {title} {\enquote
  {\bibinfo {title} {Molecular frame photoelectron angular distributions in
  dissociative photoionization of {H}$_2$ in the region of the {Q1} and {Q2}
  doubly excited states},}\ }\href {\doibase 10.1088/0953-4075/36/23/007}
  {\bibfield  {journal} {\bibinfo  {journal} {J. Phys. B: At., Mol. Opt.
  Phys.}\ }\textbf {\bibinfo {volume} {36}},\ \bibinfo {pages} {4683--4702}
  (\bibinfo {year} {2003})}\BibitemShut {NoStop}%
\bibitem [{\citenamefont {Fischer}\ \emph {et~al.}(2013)\citenamefont
  {Fischer}, \citenamefont {Sperl}, \citenamefont {C\"orlin}, \citenamefont
  {Sch\"onwald}, \citenamefont {Rietz}, \citenamefont {Palacios}, \citenamefont
  {Gonz\'alez-Castrillo}, \citenamefont {Mart\'{\i}n}, \citenamefont {Pfeifer},
  \citenamefont {Ullrich}, \citenamefont {Senftleben},\ and\ \citenamefont
  {Moshammer}}]{AIDPhysRevLett.110.213002}%
  \BibitemOpen
  \bibfield  {author} {\bibinfo {author} {\bibfnamefont {A.}~\bibnamefont
  {Fischer}}, \bibinfo {author} {\bibfnamefont {A.}~\bibnamefont {Sperl}},
  \bibinfo {author} {\bibfnamefont {P.}~\bibnamefont {C\"orlin}}, \bibinfo
  {author} {\bibfnamefont {M.}~\bibnamefont {Sch\"onwald}}, \bibinfo {author}
  {\bibfnamefont {H.}~\bibnamefont {Rietz}}, \bibinfo {author} {\bibfnamefont
  {A.}~\bibnamefont {Palacios}}, \bibinfo {author} {\bibfnamefont
  {A.}~\bibnamefont {Gonz\'alez-Castrillo}}, \bibinfo {author} {\bibfnamefont
  {F.}~\bibnamefont {Mart\'{\i}n}}, \bibinfo {author} {\bibfnamefont
  {T.}~\bibnamefont {Pfeifer}}, \bibinfo {author} {\bibfnamefont
  {J.}~\bibnamefont {Ullrich}}, \bibinfo {author} {\bibfnamefont
  {A.}~\bibnamefont {Senftleben}}, \ and\ \bibinfo {author} {\bibfnamefont
  {R.}~\bibnamefont {Moshammer}},\ }\bibfield  {title} {\enquote {\bibinfo
  {title} {Electron localization involving doubly excited states in broadband
  extreme ultraviolet ionization of {H}$_{2}$},}\ }\href {\doibase
  10.1103/PhysRevLett.110.213002} {\bibfield  {journal} {\bibinfo  {journal}
  {Phys. Rev. Lett.}\ }\textbf {\bibinfo {volume} {110}},\ \bibinfo {pages}
  {213002} (\bibinfo {year} {2013})}\BibitemShut {NoStop}%
\bibitem [{\citenamefont {Gibson}\ \emph {et~al.}(1997)\citenamefont {Gibson},
  \citenamefont {Li}, \citenamefont {Guo},\ and\ \citenamefont
  {Neira}}]{H2PhysRevLett.79.2022}%
  \BibitemOpen
  \bibfield  {author} {\bibinfo {author} {\bibfnamefont {G.~N.}\ \bibnamefont
  {Gibson}}, \bibinfo {author} {\bibfnamefont {M.}~\bibnamefont {Li}}, \bibinfo
  {author} {\bibfnamefont {C.}~\bibnamefont {Guo}}, \ and\ \bibinfo {author}
  {\bibfnamefont {J.}~\bibnamefont {Neira}},\ }\bibfield  {title} {\enquote
  {\bibinfo {title} {Strong-field dissociation and ionization of
  {H}${_{2}}^{+}$ using ultrashort laser pulses},}\ }\href {\doibase
  10.1103/PhysRevLett.79.2022} {\bibfield  {journal} {\bibinfo  {journal}
  {Phys. Rev. Lett.}\ }\textbf {\bibinfo {volume} {79}},\ \bibinfo {pages}
  {2022--2025} (\bibinfo {year} {1997})}\BibitemShut {NoStop}%
\bibitem [{\citenamefont {Lein}\ \emph {et~al.}(2002)\citenamefont {Lein},
  \citenamefont {Kreibich}, \citenamefont {Gross},\ and\ \citenamefont
  {Engel}}]{H2PhysRevA.65.033403}%
  \BibitemOpen
  \bibfield  {author} {\bibinfo {author} {\bibfnamefont {M.}~\bibnamefont
  {Lein}}, \bibinfo {author} {\bibfnamefont {T.}~\bibnamefont {Kreibich}},
  \bibinfo {author} {\bibfnamefont {E.~K.~U.}\ \bibnamefont {Gross}}, \ and\
  \bibinfo {author} {\bibfnamefont {V.}~\bibnamefont {Engel}},\ }\bibfield
  {title} {\enquote {\bibinfo {title} {Strong-field ionization dynamics of a
  model {H}$_{2}$ molecule},}\ }\href {\doibase 10.1103/PhysRevA.65.033403}
  {\bibfield  {journal} {\bibinfo  {journal} {Phys. Rev. A}\ }\textbf {\bibinfo
  {volume} {65}},\ \bibinfo {pages} {033403} (\bibinfo {year}
  {2002})}\BibitemShut {NoStop}%
\bibitem [{\citenamefont {Palacios}\ \emph {et~al.}(2015)\citenamefont
  {Palacios}, \citenamefont {Sanz-Vicario},\ and\ \citenamefont
  {Mart{\'{\i}}n}}]{H2Palacios_2015}%
  \BibitemOpen
  \bibfield  {author} {\bibinfo {author} {\bibfnamefont {A.}~\bibnamefont
  {Palacios}}, \bibinfo {author} {\bibfnamefont {J.~L.}\ \bibnamefont
  {Sanz-Vicario}}, \ and\ \bibinfo {author} {\bibfnamefont {F.}~\bibnamefont
  {Mart{\'{\i}}n}},\ }\bibfield  {title} {\enquote {\bibinfo {title}
  {Theoretical methods for attosecond electron and nuclear dynamics:
  applications to the {H}$_2$ molecule},}\ }\href {\doibase
  10.1088/0953-4075/48/24/242001} {\bibfield  {journal} {\bibinfo  {journal}
  {J. Phys. B: At., Mol. Opt. Phys.}\ }\textbf {\bibinfo {volume} {48}},\
  \bibinfo {pages} {242001} (\bibinfo {year} {2015})}\BibitemShut {NoStop}%
\bibitem [{\citenamefont {Zhang}\ \emph {et~al.}(2013)\citenamefont {Zhang},
  \citenamefont {Yuan},\ and\ \citenamefont {Zhao}}]{COPhysRevLett.111.163001}%
  \BibitemOpen
  \bibfield  {author} {\bibinfo {author} {\bibfnamefont {B.}~\bibnamefont
  {Zhang}}, \bibinfo {author} {\bibfnamefont {J.}~\bibnamefont {Yuan}}, \ and\
  \bibinfo {author} {\bibfnamefont {Z.}~\bibnamefont {Zhao}},\ }\bibfield
  {title} {\enquote {\bibinfo {title} {Dynamic core polarization in
  strong-field ionization of {CO} molecules},}\ }\href {\doibase
  10.1103/PhysRevLett.111.163001} {\bibfield  {journal} {\bibinfo  {journal}
  {Phys. Rev. Lett.}\ }\textbf {\bibinfo {volume} {111}},\ \bibinfo {pages}
  {163001} (\bibinfo {year} {2013})}\BibitemShut {NoStop}%
\bibitem [{\citenamefont {Abu-samha}\ and\ \citenamefont
  {Madsen}(2020)}]{COPhysRevA.101.013433}%
  \BibitemOpen
  \bibfield  {author} {\bibinfo {author} {\bibfnamefont {M.}~\bibnamefont
  {Abu-samha}}\ and\ \bibinfo {author} {\bibfnamefont {L.~B.}\ \bibnamefont
  {Madsen}},\ }\bibfield  {title} {\enquote {\bibinfo {title} {Effect of
  multielectron polarization in the strong-field ionization of the oriented
  {CO} molecule},}\ }\href {\doibase 10.1103/PhysRevA.101.013433} {\bibfield
  {journal} {\bibinfo  {journal} {Phys. Rev. A}\ }\textbf {\bibinfo {volume}
  {101}},\ \bibinfo {pages} {013433} (\bibinfo {year} {2020})}\BibitemShut
  {NoStop}%
\bibitem [{\citenamefont {Magrakvelidze}\ \emph {et~al.}(2012)\citenamefont
  {Magrakvelidze}, \citenamefont {Herrwerth}, \citenamefont {Jiang},
  \citenamefont {Rudenko}, \citenamefont {Kurka}, \citenamefont {Foucar},
  \citenamefont {K\"uhnel}, \citenamefont {K\"ubel}, \citenamefont {Johnson},
  \citenamefont {Schr\"oter}, \citenamefont {D\"usterer}, \citenamefont
  {Treusch}, \citenamefont {Lezius}, \citenamefont {Ben-Itzhak}, \citenamefont
  {Moshammer}, \citenamefont {Ullrich}, \citenamefont {Kling},\ and\
  \citenamefont {Thumm}}]{N2PhysRevA.86.013415}%
  \BibitemOpen
  \bibfield  {author} {\bibinfo {author} {\bibfnamefont {M.}~\bibnamefont
  {Magrakvelidze}}, \bibinfo {author} {\bibfnamefont {O.}~\bibnamefont
  {Herrwerth}}, \bibinfo {author} {\bibfnamefont {Y.~H.}\ \bibnamefont
  {Jiang}}, \bibinfo {author} {\bibfnamefont {A.}~\bibnamefont {Rudenko}},
  \bibinfo {author} {\bibfnamefont {M.}~\bibnamefont {Kurka}}, \bibinfo
  {author} {\bibfnamefont {L.}~\bibnamefont {Foucar}}, \bibinfo {author}
  {\bibfnamefont {K.~U.}\ \bibnamefont {K\"uhnel}}, \bibinfo {author}
  {\bibfnamefont {M.}~\bibnamefont {K\"ubel}}, \bibinfo {author} {\bibfnamefont
  {N.~G.}\ \bibnamefont {Johnson}}, \bibinfo {author} {\bibfnamefont {C.~D.}\
  \bibnamefont {Schr\"oter}}, \bibinfo {author} {\bibfnamefont
  {S.}~\bibnamefont {D\"usterer}}, \bibinfo {author} {\bibfnamefont
  {R.}~\bibnamefont {Treusch}}, \bibinfo {author} {\bibfnamefont
  {M.}~\bibnamefont {Lezius}}, \bibinfo {author} {\bibfnamefont
  {I.}~\bibnamefont {Ben-Itzhak}}, \bibinfo {author} {\bibfnamefont
  {R.}~\bibnamefont {Moshammer}}, \bibinfo {author} {\bibfnamefont
  {J.}~\bibnamefont {Ullrich}}, \bibinfo {author} {\bibfnamefont {M.~F.}\
  \bibnamefont {Kling}}, \ and\ \bibinfo {author} {\bibfnamefont
  {U.}~\bibnamefont {Thumm}},\ }\bibfield  {title} {\enquote {\bibinfo {title}
  {Tracing nuclear-wave-packet dynamics in singly and doubly charged states of
  {N}${}_{2}$ and {O}${}_{2}$ with {XUV}-pump--{XUV}-probe experiments},}\
  }\href {\doibase 10.1103/PhysRevA.86.013415} {\bibfield  {journal} {\bibinfo
  {journal} {Phys. Rev. A}\ }\textbf {\bibinfo {volume} {86}},\ \bibinfo
  {pages} {013415} (\bibinfo {year} {2012})}\BibitemShut {NoStop}%
\bibitem [{\citenamefont {Lehmann}\ \emph {et~al.}(2016)\citenamefont
  {Lehmann}, \citenamefont {Pic\'on}, \citenamefont {Bostedt}, \citenamefont
  {Rudenko}, \citenamefont {Marinelli}, \citenamefont {Moonshiram},
  \citenamefont {Osipov}, \citenamefont {Rolles}, \citenamefont {Berrah},
  \citenamefont {Bomme}, \citenamefont {Bucher}, \citenamefont {Doumy},
  \citenamefont {Erk}, \citenamefont {Ferguson}, \citenamefont {Gorkhover},
  \citenamefont {Ho}, \citenamefont {Kanter}, \citenamefont {Kr\"assig},
  \citenamefont {Krzywinski}, \citenamefont {Lutman}, \citenamefont {March},
  \citenamefont {Ray}, \citenamefont {Young}, \citenamefont {Pratt},\ and\
  \citenamefont {Southworth}}]{N2PhysRevA.94.013426}%
  \BibitemOpen
  \bibfield  {author} {\bibinfo {author} {\bibfnamefont {C.~S.}\ \bibnamefont
  {Lehmann}}, \bibinfo {author} {\bibfnamefont {A.}~\bibnamefont {Pic\'on}},
  \bibinfo {author} {\bibfnamefont {C.}~\bibnamefont {Bostedt}}, \bibinfo
  {author} {\bibfnamefont {A.}~\bibnamefont {Rudenko}}, \bibinfo {author}
  {\bibfnamefont {A.}~\bibnamefont {Marinelli}}, \bibinfo {author}
  {\bibfnamefont {D.}~\bibnamefont {Moonshiram}}, \bibinfo {author}
  {\bibfnamefont {T.}~\bibnamefont {Osipov}}, \bibinfo {author} {\bibfnamefont
  {D.}~\bibnamefont {Rolles}}, \bibinfo {author} {\bibfnamefont
  {N.}~\bibnamefont {Berrah}}, \bibinfo {author} {\bibfnamefont
  {C.}~\bibnamefont {Bomme}}, \bibinfo {author} {\bibfnamefont
  {M.}~\bibnamefont {Bucher}}, \bibinfo {author} {\bibfnamefont
  {G.}~\bibnamefont {Doumy}}, \bibinfo {author} {\bibfnamefont
  {B.}~\bibnamefont {Erk}}, \bibinfo {author} {\bibfnamefont {K.~R.}\
  \bibnamefont {Ferguson}}, \bibinfo {author} {\bibfnamefont {T.}~\bibnamefont
  {Gorkhover}}, \bibinfo {author} {\bibfnamefont {P.~J.}\ \bibnamefont {Ho}},
  \bibinfo {author} {\bibfnamefont {E.~P.}\ \bibnamefont {Kanter}}, \bibinfo
  {author} {\bibfnamefont {B.}~\bibnamefont {Kr\"assig}}, \bibinfo {author}
  {\bibfnamefont {J.}~\bibnamefont {Krzywinski}}, \bibinfo {author}
  {\bibfnamefont {A.~A.}\ \bibnamefont {Lutman}}, \bibinfo {author}
  {\bibfnamefont {A.~M.}\ \bibnamefont {March}}, \bibinfo {author}
  {\bibfnamefont {D.}~\bibnamefont {Ray}}, \bibinfo {author} {\bibfnamefont
  {L.}~\bibnamefont {Young}}, \bibinfo {author} {\bibfnamefont {S.~T.}\
  \bibnamefont {Pratt}}, \ and\ \bibinfo {author} {\bibfnamefont {S.~H.}\
  \bibnamefont {Southworth}},\ }\bibfield  {title} {\enquote {\bibinfo {title}
  {Ultrafast x-ray-induced nuclear dynamics in diatomic molecules using
  femtosecond x-ray-pump--x-ray-probe spectroscopy},}\ }\href {\doibase
  10.1103/PhysRevA.94.013426} {\bibfield  {journal} {\bibinfo  {journal} {Phys.
  Rev. A}\ }\textbf {\bibinfo {volume} {94}},\ \bibinfo {pages} {013426}
  (\bibinfo {year} {2016})}\BibitemShut {NoStop}%
\bibitem [{\citenamefont {Guo}\ \emph {et~al.}(1998)\citenamefont {Guo},
  \citenamefont {Li}, \citenamefont {Nibarger},\ and\ \citenamefont
  {Gibson}}]{N2CPhysRevA.58.R4271}%
  \BibitemOpen
  \bibfield  {author} {\bibinfo {author} {\bibfnamefont {C.}~\bibnamefont
  {Guo}}, \bibinfo {author} {\bibfnamefont {M.}~\bibnamefont {Li}}, \bibinfo
  {author} {\bibfnamefont {J.~P.}\ \bibnamefont {Nibarger}}, \ and\ \bibinfo
  {author} {\bibfnamefont {G.~N.}\ \bibnamefont {Gibson}},\ }\bibfield  {title}
  {\enquote {\bibinfo {title} {Single and double ionization of diatomic
  molecules in strong laser fields},}\ }\href {\doibase
  10.1103/PhysRevA.58.R4271} {\bibfield  {journal} {\bibinfo  {journal} {Phys.
  Rev. A}\ }\textbf {\bibinfo {volume} {58}},\ \bibinfo {pages} {R4271--R4274}
  (\bibinfo {year} {1998})}\BibitemShut {NoStop}%
\bibitem [{\citenamefont {Guo}\ and\ \citenamefont
  {Gibson}(2001)}]{N2CPhysRevA.63.040701}%
  \BibitemOpen
  \bibfield  {author} {\bibinfo {author} {\bibfnamefont {C.}~\bibnamefont
  {Guo}}\ and\ \bibinfo {author} {\bibfnamefont {G.~N.}\ \bibnamefont
  {Gibson}},\ }\bibfield  {title} {\enquote {\bibinfo {title} {Ellipticity
  effects on single and double ionization of diatomic molecules in strong laser
  fields},}\ }\href {\doibase 10.1103/PhysRevA.63.040701} {\bibfield  {journal}
  {\bibinfo  {journal} {Phys. Rev. A}\ }\textbf {\bibinfo {volume} {63}},\
  \bibinfo {pages} {040701} (\bibinfo {year} {2001})}\BibitemShut {NoStop}%
\bibitem [{\citenamefont {Eremina}\ \emph {et~al.}(2004)\citenamefont
  {Eremina}, \citenamefont {Liu}, \citenamefont {Rottke}, \citenamefont
  {Sandner}, \citenamefont {Sch\"atzel}, \citenamefont {Dreischuh},
  \citenamefont {Paulus}, \citenamefont {Walther}, \citenamefont {Moshammer},\
  and\ \citenamefont {Ullrich}}]{N2CPhysRevLett.92.173001}%
  \BibitemOpen
  \bibfield  {author} {\bibinfo {author} {\bibfnamefont {E.}~\bibnamefont
  {Eremina}}, \bibinfo {author} {\bibfnamefont {X.}~\bibnamefont {Liu}},
  \bibinfo {author} {\bibfnamefont {H.}~\bibnamefont {Rottke}}, \bibinfo
  {author} {\bibfnamefont {W.}~\bibnamefont {Sandner}}, \bibinfo {author}
  {\bibfnamefont {M.~G.}\ \bibnamefont {Sch\"atzel}}, \bibinfo {author}
  {\bibfnamefont {A.}~\bibnamefont {Dreischuh}}, \bibinfo {author}
  {\bibfnamefont {G.~G.}\ \bibnamefont {Paulus}}, \bibinfo {author}
  {\bibfnamefont {H.}~\bibnamefont {Walther}}, \bibinfo {author} {\bibfnamefont
  {R.}~\bibnamefont {Moshammer}}, \ and\ \bibinfo {author} {\bibfnamefont
  {J.}~\bibnamefont {Ullrich}},\ }\bibfield  {title} {\enquote {\bibinfo
  {title} {Influence of molecular structure on double ionization of {N}$_{2}$
  and {O}$_{2}$ by high intensity ultrashort laser pulses},}\ }\href {\doibase
  10.1103/PhysRevLett.92.173001} {\bibfield  {journal} {\bibinfo  {journal}
  {Phys. Rev. Lett.}\ }\textbf {\bibinfo {volume} {92}},\ \bibinfo {pages}
  {173001} (\bibinfo {year} {2004})}\BibitemShut {NoStop}%
\bibitem [{\citenamefont {Bocharova}\ \emph
  {et~al.}(2011{\natexlab{b}})\citenamefont {Bocharova}, \citenamefont
  {Alnaser}, \citenamefont {Thumm}, \citenamefont {Niederhausen}, \citenamefont
  {Ray}, \citenamefont {Cocke},\ and\ \citenamefont
  {Litvinyuk}}]{N2insPhysRevA.83.013417}%
  \BibitemOpen
  \bibfield  {author} {\bibinfo {author} {\bibfnamefont {I.~A.}\ \bibnamefont
  {Bocharova}}, \bibinfo {author} {\bibfnamefont {A.~S.}\ \bibnamefont
  {Alnaser}}, \bibinfo {author} {\bibfnamefont {U.}~\bibnamefont {Thumm}},
  \bibinfo {author} {\bibfnamefont {T.}~\bibnamefont {Niederhausen}}, \bibinfo
  {author} {\bibfnamefont {D.}~\bibnamefont {Ray}}, \bibinfo {author}
  {\bibfnamefont {C.~L.}\ \bibnamefont {Cocke}}, \ and\ \bibinfo {author}
  {\bibfnamefont {I.~V.}\ \bibnamefont {Litvinyuk}},\ }\bibfield  {title}
  {\enquote {\bibinfo {title} {Time-resolved {C}oulomb-explosion imaging of
  nuclear wave-packet dynamics induced in diatomic molecules by intense
  few-cycle laser pulses},}\ }\href {\doibase 10.1103/PhysRevA.83.013417}
  {\bibfield  {journal} {\bibinfo  {journal} {Phys. Rev. A}\ }\textbf {\bibinfo
  {volume} {83}},\ \bibinfo {pages} {013417} (\bibinfo {year}
  {2011}{\natexlab{b}})}\BibitemShut {NoStop}%
\bibitem [{\citenamefont {Lucchini}\ \emph {et~al.}(2012)\citenamefont
  {Lucchini}, \citenamefont {Kim}, \citenamefont {Calegari}, \citenamefont
  {Kelkensberg}, \citenamefont {Siu}, \citenamefont {Sansone}, \citenamefont
  {Vrakking}, \citenamefont {Hochlaf},\ and\ \citenamefont
  {Nisoli}}]{N2insPhysRevA.86.043404}%
  \BibitemOpen
  \bibfield  {author} {\bibinfo {author} {\bibfnamefont {M.}~\bibnamefont
  {Lucchini}}, \bibinfo {author} {\bibfnamefont {K.}~\bibnamefont {Kim}},
  \bibinfo {author} {\bibfnamefont {F.}~\bibnamefont {Calegari}}, \bibinfo
  {author} {\bibfnamefont {F.}~\bibnamefont {Kelkensberg}}, \bibinfo {author}
  {\bibfnamefont {W.}~\bibnamefont {Siu}}, \bibinfo {author} {\bibfnamefont
  {G.}~\bibnamefont {Sansone}}, \bibinfo {author} {\bibfnamefont {M.~J.~J.}\
  \bibnamefont {Vrakking}}, \bibinfo {author} {\bibfnamefont {M.}~\bibnamefont
  {Hochlaf}}, \ and\ \bibinfo {author} {\bibfnamefont {M.}~\bibnamefont
  {Nisoli}},\ }\bibfield  {title} {\enquote {\bibinfo {title} {Autoionization
  and ultrafast relaxation dynamics of highly excited states in {N}$_{2}$},}\
  }\href {\doibase 10.1103/PhysRevA.86.043404} {\bibfield  {journal} {\bibinfo
  {journal} {Phys. Rev. A}\ }\textbf {\bibinfo {volume} {86}},\ \bibinfo
  {pages} {043404} (\bibinfo {year} {2012})}\BibitemShut {NoStop}%
\bibitem [{\citenamefont {Hanna}\ \emph {et~al.}(2017)\citenamefont {Hanna},
  \citenamefont {Vendrell}, \citenamefont {Ourmazd},\ and\ \citenamefont
  {Santra}}]{N2insPhysRevA.95.043419}%
  \BibitemOpen
  \bibfield  {author} {\bibinfo {author} {\bibfnamefont {A.~M.}\ \bibnamefont
  {Hanna}}, \bibinfo {author} {\bibfnamefont {O.}~\bibnamefont {Vendrell}},
  \bibinfo {author} {\bibfnamefont {A.}~\bibnamefont {Ourmazd}}, \ and\
  \bibinfo {author} {\bibfnamefont {R.}~\bibnamefont {Santra}},\ }\bibfield
  {title} {\enquote {\bibinfo {title} {Laser control over the ultrafast coulomb
  explosion of {N}$_{2}^{2+}$ after auger decay: A quantum-dynamics
  investigation},}\ }\href {\doibase 10.1103/PhysRevA.95.043419} {\bibfield
  {journal} {\bibinfo  {journal} {Phys. Rev. A}\ }\textbf {\bibinfo {volume}
  {95}},\ \bibinfo {pages} {043419} (\bibinfo {year} {2017})}\BibitemShut
  {NoStop}%
\bibitem [{\citenamefont {Leth}\ \emph
  {et~al.}(2010{\natexlab{a}})\citenamefont {Leth}, \citenamefont {Madsen},\
  and\ \citenamefont {M\o{}lmer}}]{MCWPH2PhysRevA.81.053409}%
  \BibitemOpen
  \bibfield  {author} {\bibinfo {author} {\bibfnamefont {H.~A.}\ \bibnamefont
  {Leth}}, \bibinfo {author} {\bibfnamefont {L.~B.}\ \bibnamefont {Madsen}}, \
  and\ \bibinfo {author} {\bibfnamefont {K.}~\bibnamefont {M\o{}lmer}},\
  }\bibfield  {title} {\enquote {\bibinfo {title} {Monte carlo wave packet
  approach to dissociative multiple ionization in diatomic molecules},}\ }\href
  {\doibase 10.1103/PhysRevA.81.053409} {\bibfield  {journal} {\bibinfo
  {journal} {Phys. Rev. A}\ }\textbf {\bibinfo {volume} {81}},\ \bibinfo
  {pages} {053409} (\bibinfo {year} {2010}{\natexlab{a}})}\BibitemShut
  {NoStop}%
\bibitem [{\citenamefont {Leth}\ \emph
  {et~al.}(2010{\natexlab{b}})\citenamefont {Leth}, \citenamefont {Madsen},\
  and\ \citenamefont {M\o{}lmer}}]{MCWPH2PhysRevA.81.053410}%
  \BibitemOpen
  \bibfield  {author} {\bibinfo {author} {\bibfnamefont {H.~A.}\ \bibnamefont
  {Leth}}, \bibinfo {author} {\bibfnamefont {L.~B.}\ \bibnamefont {Madsen}}, \
  and\ \bibinfo {author} {\bibfnamefont {K.}~\bibnamefont {M\o{}lmer}},\
  }\bibfield  {title} {\enquote {\bibinfo {title} {Dissociative double
  ionization of {H}$_{2}$ and {D}$_{2}$: Comparison between experiment and
  {M}onte {C}arlo wave packet calculations},}\ }\href {\doibase
  10.1103/PhysRevA.81.053410} {\bibfield  {journal} {\bibinfo  {journal} {Phys.
  Rev. A}\ }\textbf {\bibinfo {volume} {81}},\ \bibinfo {pages} {053410}
  (\bibinfo {year} {2010}{\natexlab{b}})}\BibitemShut {NoStop}%
\bibitem [{\citenamefont {Jing}\ and\ \citenamefont
  {Madsen}(2016)}]{MCWPH2PhysRevA.94.063402}%
  \BibitemOpen
  \bibfield  {author} {\bibinfo {author} {\bibfnamefont {Q.}~\bibnamefont
  {Jing}}\ and\ \bibinfo {author} {\bibfnamefont {L.~B.}\ \bibnamefont
  {Madsen}},\ }\bibfield  {title} {\enquote {\bibinfo {title} {Laser-induced
  dissociative ionization of {H}$_{2}$ from the near-infrared to the
  mid-infrared regime},}\ }\href {\doibase 10.1103/PhysRevA.94.063402}
  {\bibfield  {journal} {\bibinfo  {journal} {Phys. Rev. A}\ }\textbf {\bibinfo
  {volume} {94}},\ \bibinfo {pages} {063402} (\bibinfo {year}
  {2016})}\BibitemShut {NoStop}%
\bibitem [{\citenamefont {Jing}\ \emph {et~al.}(2018)\citenamefont {Jing},
  \citenamefont {Bello}, \citenamefont {Mart\'{\i}n}, \citenamefont
  {Palacios},\ and\ \citenamefont {Madsen}}]{MCWPH2PhysRevA.97.043426}%
  \BibitemOpen
  \bibfield  {author} {\bibinfo {author} {\bibfnamefont {Q.}~\bibnamefont
  {Jing}}, \bibinfo {author} {\bibfnamefont {R.~Y.}\ \bibnamefont {Bello}},
  \bibinfo {author} {\bibfnamefont {F.}~\bibnamefont {Mart\'{\i}n}}, \bibinfo
  {author} {\bibfnamefont {A.}~\bibnamefont {Palacios}}, \ and\ \bibinfo
  {author} {\bibfnamefont {L.~B.}\ \bibnamefont {Madsen}},\ }\bibfield  {title}
  {\enquote {\bibinfo {title} {Monte carlo wave-packet approach to trace
  nuclear dynamics in molecular excited states by {XUV}-pump--{IR}-probe
  spectroscopy},}\ }\href {\doibase 10.1103/PhysRevA.97.043426} {\bibfield
  {journal} {\bibinfo  {journal} {Phys. Rev. A}\ }\textbf {\bibinfo {volume}
  {97}},\ \bibinfo {pages} {043426} (\bibinfo {year} {2018})}\BibitemShut
  {NoStop}%
\bibitem [{\citenamefont {Leth}\ and\ \citenamefont
  {Madsen}(2011)}]{MCWPO2PhysRevA.83.063415}%
  \BibitemOpen
  \bibfield  {author} {\bibinfo {author} {\bibfnamefont {H.~A.}\ \bibnamefont
  {Leth}}\ and\ \bibinfo {author} {\bibfnamefont {L.~B.}\ \bibnamefont
  {Madsen}},\ }\bibfield  {title} {\enquote {\bibinfo {title} {Dissociative
  multiple ionization of diatomic molecules by extreme-ultraviolet
  free-electron-laser pulses},}\ }\href {\doibase 10.1103/PhysRevA.83.063415}
  {\bibfield  {journal} {\bibinfo  {journal} {Phys. Rev. A}\ }\textbf {\bibinfo
  {volume} {83}},\ \bibinfo {pages} {063415} (\bibinfo {year}
  {2011})}\BibitemShut {NoStop}%
\bibitem [{\citenamefont {Feit}\ \emph {et~al.}(1982)\citenamefont {Feit},
  \citenamefont {Fleck},\ and\ \citenamefont {Steiger}}]{FFTFEIT1982412}%
  \BibitemOpen
  \bibfield  {author} {\bibinfo {author} {\bibfnamefont {M.}~\bibnamefont
  {Feit}}, \bibinfo {author} {\bibfnamefont {J.}~\bibnamefont {Fleck}}, \ and\
  \bibinfo {author} {\bibfnamefont {A.}~\bibnamefont {Steiger}},\ }\bibfield
  {title} {\enquote {\bibinfo {title} {Solution of the schr\"{o}dinger equation
  by a spectral method},}\ }\href {\doibase
  https://doi.org/10.1016/0021-9991(82)90091-2} {\bibfield  {journal} {\bibinfo
   {journal} {J. Comput. Phys.}\ }\textbf {\bibinfo {volume} {47}},\ \bibinfo
  {pages} {412--433} (\bibinfo {year} {1982})}\BibitemShut {NoStop}%
\end{thebibliography}%
%
%
\end{document}